\documentclass[a4paper,fleqn,usenatbib]{mnras}

\usepackage{newtxtext,newtxmath}

\usepackage[T1]{fontenc}
\usepackage{ae,aecompl,multirow}

\usepackage{graphicx}	
\usepackage{amsmath}	
\usepackage{amssymb}	
\usepackage{subfig}

\title[Unfolding the X-ray spectral curvature of Mkn\,421]{Unfolding the X-ray Spectral Curvature of Mkn\,421 for Further Clues}

\author[P. Goswami et al.]{
Pranjupriya Goswami$^{1}$\thanks{E-mail: pranjupriya.g@gmail.com},
Sunder Sahayanathan$^{2,3}$ \thanks{Email: sunder@barc.gov.in},
Atreyee Sinha$^{4}$ 
$\&$ Rupjyoti Gogoi$^{1}$
\\
$^{1}$Department of Physics, Tezpur University, Napaam - 784028, India\\
$^{2}$Astrophysical Sciences Division, Bhabha Atomic Research Centre, Mumbai - 400085, India\\
$^{3}$Homi Bhabha National Institute, Mumbai - 400094, India\\
$^{4}$Laboratoire  Univers  et  Particules  de  Montpellier, Universit\'e de Montpellier, CNRS, Montpellier - 34095, France.
}

\date{Accepted XXX. Received YYY; in original form ZZZ}

\pubyear{2020}

\begin{document}
\label{firstpage}
\pagerange{\pageref{firstpage}--\pageref{lastpage}}
\maketitle

\begin{abstract}

The X-ray observations of Mkn\,421 show significant spectral curvature that can be reproduced by a log-parabola function.
The spectra can also be fitted by an analytical model considering synchrotron emission from an electron distribution that 
is accelerated at a shock front with an energy-dependent diffusion(EDD model). The spectral fit of \emph{Nu}STAR and 
\emph{Swift}-XRT observations using EDD model during different flux states reveal the model parameters are strongly 
correlated. We perform a detailed investigation of this correlation to decipher the information hidden underneath. 
The model predicts the synchrotron peak energy to be correlated with the peak spectral curvature which is consistent 
with the case of Mkn\,421. Expressing the energy dependence of the diffusion in terms of the magnetohydrodynamic turbulence 
energy index, it appears the turbulence shifts from Kolmogorov/Kraichnan type to Bohm limit during high flux states. 
Further, the correlation between the best-fit parameters of EDD model lets us derive an expression for the product of source magnetic 
field($B$) and jet Doppler factor($\delta$) in terms of synchrotron and Compton peak energies. 
The synchrotron peak energy is obtained using the simultaneous \emph{Swift}-XRT--\emph{Nu}STAR observations;
whereas, the Compton peak energy is estimated by performing a linear regression analysis of the archival spectral peaks.
The deduced $\delta B$  varies over a wide range; however, it satisfies reasonably well with the values estimated 
solely from the spectral peak energies independent of the EDD model.
This highlights the plausible connection between the microscopic description of the electron diffusion with the 
macroscopic quantities deciding the broadband spectrum of Mkn\,421.

\end{abstract}


\section{Introduction}
Mkn\,421 is one of the nearest BL Lac type of active galactic nuclei located at redshift $z=0.031$ \citep{Punch+1992}.
The spectra of these class of sources extend from radio to GeV/TeV energies and also found to be rapidly varying in a time-scale 
of days to few minutes \citep{1997ICRC....3..257M,Tanihata+2004,2009A&A...501..879T,Donnarumma+2009,Aleksic+2015,Hovatta+2015,Sinha2015}. 
The detection of very high energy radiation and the rapid variability confirm the emission is significantly Doppler boosted due to the 
relativistic motion of the emitting region \citep{1995MNRAS.273..583D}. Their spectral energy distribution (SED) is characterized by two 
prominent peaks with the low energy component peaking at UV/X-ray energy and well understood to be synchrotron emission from a non-thermal 
electron distribution \citep{Padovani+1995}. The high energy component peaking at $\gamma$-ray energy is usually attributed to synchrotron 
self Compton emission \citep{Abdo+2011, Acciari+2011}. In case of Mkn\,421, this interpretation of high energy emission process is further 
supported by the correlated X-ray/TeV flares and the quadratic dependence of the X-ray and $\gamma$-ray fractional variability 
\citep{1997A&A...320...19M,2005ApJ...630..130B,Giebels+2007,Fossati+2008,2010A&A...510A..63K,2011ApJ...734..110B,2018ApJ...854...66K,2018ApJ...858...68K}. 
Interestingly, its synchrotron peak is observed to vary over a broad energy range during different flux states 
\citep{Aleksic+2015b, Kapanadze+2016, bartoli2016, mislav2016}.

The narrow band X-ray spectra of Mkn\,421 exhibits significant curvature that cannot be interpreted by a simple power-law or a power-law 
with an exponential cutoff \citep{2000ApJ...541..166F,2001ApJ...559..187K, 2004A&A...413..489M,2007A&A...466..521T,2013ApJ...765..122Y}. 
The curvature extends even at hard X-rays and hence cannot be easily attributed to the one arising from the spectral changeover at the peak 
of the synchrotron component \citep{Fossati+2008, 2009ApJ...695..596H}. On the other hand, a log-parabola function is able to explain the 
X-ray spectral curvature of Mkn\,421 over a narrow band; however, it fails to reproduce the optical/UV flux when extended to low energies 
\citep{2004A&A...413..489M,Sinha2015}. 

A log-parabola spectrum demands the underlying electron distribution also to be a log-parabola and such an electron distribution can be attained 
when the particle acceleration probability is energy-dependent \citep{2004A&A...413..489M}. The probability for confining the particle within the 
acceleration region will decrease with the increase in the gyroradius and hence, the acceleration probability can be energy-dependent.
This model predicts the synchrotron peak energy to be anti-correlated with its curvature \citep{2007A&A...466..521T}. Further using Monte Carlo 
simulations, \cite{2011ApJ...739...66T} showed the electron distribution resulting from a stochastic acceleration can be approximated by a 
log-parabola function. Using this time-dependent model, the authors studied the evolution of the accelerated electron distribution under different 
types of turbulent energy spectrum. They showed the curvature at the peak energy of the accelerated electron distribution is related to 
the acceleration time-scale associated with the momentum diffusion. Subsequently, they predicted the peak spectral curvature of the synchrotron 
spectrum to be anti-correlated with the peak energy. Analysis of the log-parabola X-ray spectral shape of Mkn\,421 and BL Lacs 
in general, are largely inconclusive. Modelling the X-ray observation of Mkn\,421 and other TeV BL Lacs, during 1997-2006,
using a log-parabola model have revealed an anti-correlation between the peak spectral curvature and the peak energy
consistent with the stochastic acceleration process \citep{2007A&A...466..521T,2008A&A...478..395M,2009A&A...501..879T, 2011ApJ...739...66T}. 
On the contrary, X-ray analysis for Mkn\,421 using \emph{Swift}-XRT/\emph{Nu}STAR data during 2005-2008 and during 2009-2013
 \citep{Sinha2015,Kapanadze+2016,2018ApJ...854...66K,2018ApJ...858...68K} have shown weak negative/no correlation between these quantities. 

Time-dependent models, tracing the evolution of the accelerated particle distribution and the emission spectra, under energy-dependent acceleration 
and/or escape time-scales have been studied by several authors, and are capable of producing variety of spectra 
\citep{Stawarz+2008,Lewis+2016,2014MNRAS.442.3166Z,Summerlin_2011,Baring+2017,Kakuwa+2016,2014ApJ...780...64A,Becker+2006,Katarzynski+2006}. 
For stochastic acceleration process, the energy-dependence of the acceleration and escape time-scale can be expressed as $t_{\rm acc}\sim E^{2-q}$ 
and $t_{\rm esc}\sim E^{q-2}$ where, $E$ is the particle energy and $q$ is the power-law index of the magnetohydrodynamic turbulent spectrum 
\citep{Stawarz+2008,2011ApJ...739...66T}. Additionally, in case of shock acceleration, $t_{\rm acc}$ is also determined by the time-scale at which 
particles cycle across the shock \citep{Lewis+2016}. The value of $q$ can range from $q=1$ associated with Bohm diffusion, to $q=2$ under 
hard-sphere approximation \citep{1998A&A...333..452K, 2014ApJ...780...64A}. In case of Kolmogorov type turbulence $q=5/3$ and for Kraichnan type 
turbulence $q=3/2$ \citep{2019ApJ...877...71T}.
Often these time-dependent models involve large number of parameters and are mainly used to understand the observations from a theoretical perspective. 
Instead, reducing the accelerated particle distribution and the emission spectrum into a simple mathematical form (similar to a log-parabola function) 
can help us to perform a detailed statistical analysis of the observed data that provide a deeper insight into the 
physics of the source \citep[e.g.][]{2004A&A...413..489M,2011ApJ...739...66T,pgoswami2018,Sitha2018}.

Under the simplistic approach, a curved spectrum can also be interpreted as an outcome of the energy-dependent diffusion from the acceleration region 
(EDD model). The synchrotron spectral shape due to this model is determined by two parameters and hence a direct comparison with the log-parabola function 
is possible. In our earlier work, we demonstrate the capability of this model to fit the \emph{Nu}STAR and \emph{Swift}-XRT X-ray observations of Mkn\,421 
and the spectral fits are comparable with that of a log-parabola (\citealp{pgoswami2018}, hereafter Paper-I). Unlike the later, the best-fit parameters are well 
correlated in the case of EDD model and the correlation is much stronger when the fitting is performed on the \emph{Nu}STAR data alone.\\

In this work, we revisit the EDD model to investigate the correlation between the fit parameters and interpret the physical conditions of Mkn\,421 
during various flux states. The model is also upgraded by considering the finite width of the synchrotron single particle emissivity rather than a 
$\delta$-function approximation and the total emissivity is estimated numerically. Further, we study the salient features of the observed 
spectrum that can be verified through observations. The modified EDD model is applied to \emph{Nu}STAR and \emph{Swift}-XRT X-ray observations of 
Mkn\,421 and the best-fit parameters are obtained. In addition, we also update our study by including the recent X-ray observations. 
The energy-dependence of the diffusive process can represent the magnetohydrodynamic turbulence in 
blazar jet and hence, the EDD model can provide a convenient way to probe the same. Finally, we show that the correlation between the best-fit EDD 
parameters can be used to establish a relation between the energy dependence of the electron escape time-scale, and the product of the source magnetic 
field and the jet Doppler factor. This illustrates that the information regarding the physical parameters deciding the broadband emission of 
the source is concealed within the X-ray spectral shape. In the following section, we describe the formalism of EDD model in detail and observational 
features of the synchrotron spectrum. In \S \ref{sec:analysis}, we describe the analysis of the X-ray spectra of Mkn\,421 and present the correlation 
study results. Following this in \S \ref{sec:srcchar}, we show how the correlation between the best-fit parameters can be used to obtain a relation 
between the energy dependence of the diffusive process and the source parameters, and extend this for the case of Mkn\,421. 

\section{Energy-dependent Diffusion (EDD) Model}\label{sec:eed}

We consider the non-thermal electron distribution $N_a(E)$, responsible for the broadband emission in blazars,
to be accelerated at a shock front through first-order Fermi mechanism and its evolution is governed by \citep{1962SvA.....6..317K,1998A&A...333..452K, 2000ApJ...536..299K}
\begin{align}\label{eq:ke}
	\frac{\partial\,N_a(E)}{\partial t}+\frac{\partial}{\partial E}\left[\left(\frac{E}{t_{\rm acc}}-\beta_sE^2\right) N_a(E,t)\right]=& Q_0 \delta(E-E_{\rm inj}) \nonumber \\ &- \frac{N_a(E,t)}{t_{\rm esc}}
\end{align}
Here, $E$ is the electron energy and the particle acceleration rate at the shock front is parameterized as $1/t_{\rm acc}$ while 
the radiative loss rate due to synchrotron emission is given as $\beta_sE^2$. $\beta_s$ is related to the magnetic energy density of the region in vicinity to the shock front \citep{1998A&A...333..452K,Rybicki_Lightman}. We consider the case, where the electrons gain energy mainly by crossing the shock front 
and the change in energy by individual scatterings at magnetic inhomogeneities are neglected. Under this scenario, 
$t_{\rm acc}$ will be determined by the time-scale at which particles cycle across the shock, and if we assume the spatial
diffusion coefficient in the vicinity of the shock to be constant then, $t_{\rm acc}=$ constant \citep{Lewis+2016}.
Further, we assume a monoenergetic injection of electron at energy $E_{\rm inj}$ and 
the accelerated electron escape from the acceleration region (vicinity of the shock front) at a rate $1/t_{\rm esc}$. 
The magnetohydrodynamic turbulence in jet flow can cause the spatial diffusion of the electron to be energy-dependent and
we imitate this by considering the electron escape time-scale from acceleration region to be of the form,

\begin{align}\label{eq:tesc}
	t_{\rm esc}(E) \propto \left(\frac{E}{E_0}\right)^{-\kappa} 
\end{align}
where, $\kappa$ determines the energy dependence of the electron escape and the proportionality constant is 
the electron escape time-scale corresponding to electron energy $E_0$.
Expressing the radiative loss in terms of maximum accelerated electron 
energy $E_{\rm max} = (\beta_s\, t_{\rm acc})^{-1}$ \citep{1998A&A...333..452K},
the steady state condition of equation (\ref{eq:ke}) will be
\begin{align}\label{eq:kin}
	\frac{d}{d\, E}\left[\left(1-\frac{E}{E_{\rm max}}\right)\frac{E}{t_{\rm acc}} N_a(E)\right]= Q_0 \delta(E-E_{\rm inj})- \frac{N_a(E)}{t_{\rm esc}}
\end{align}

Under constant $t_{\rm acc}$ and for $E_{\rm inj}< E \ll E_{\rm max}$, equation (\ref{eq:kin}) can be reduced to (\citealp{Sitha2018}, Paper-I). 

\begin{align}\label{eq:eq2}
	\frac{d\, {\rm ln}[E N_a(E)]}{d\, {\rm ln}E} \approx - \frac{t_{\rm acc}}{t_{\rm esc}}
\end{align}
Using equation (\ref{eq:tesc}), the solution of above will be

\begin{align}\label{eq:na}
	N_a(E) \propto E^{-1} {\rm exp}\left[-\frac {\xi_0}{\kappa}\left(\frac{E}{E_0}\right)^\kappa \right]
\end{align}
Here, $\xi_0$ is the ratio $t_{\rm acc}/t_{\rm esc}$ at energy $E_0$. 
The accelerated electrons will eventually escape from the acceleration region and lose their energy through 
synchrotron and inverse Compton emission processes. Then the 
steady state equation governing the electron distribution $N_c$ responsible for the observed emission spectrum 
will be \citep{1962SvA.....6..317K,1998A&A...333..452K, 2000ApJ...536..299K},
\begin{align}
	-\frac{d}{d\, E}\left[B E^2 N_c(E)\right]= \frac{N_a(E)}{t_{\rm esc}}
\end{align}
where, $BE^2$ is the radiative energy loss rate and the right-hand side is the injection from the acceleration region. 
Using equations (\ref{eq:tesc}) and (\ref{eq:na}) we get,

\begin{align}\label{eq:nce}
	N_c(E) \propto E^{-2} {\rm exp}\left[-\frac {\xi_0}{\kappa}\left(\frac{E}{E_0}\right)^\kappa \right]
\end{align}

The observed synchrotron flux $F_\epsilon$ at energy $\epsilon$ can be attained by convolving $N_c(E)$ 
with the single particle emissivity function after accounting for the effects due to relativistic motion of the emission 
region and cosmology \citep{1984RvMP...56..255B,Rybicki_Lightman}:
\begin{align}\label{eq:synf}
	F_\epsilon &\propto \int\limits_0^\infty f\left(\frac{1+z}{\delta} \frac {\epsilon}{\epsilon_c}\right) N_c(E)\, dE {\rm ;} \\
	\epsilon_c &= \frac{3heB}{16m^2c^3}\left(\frac{E^2}{mc^2}\right)
\end{align}
Here, $z$ is the redshift of the source, $\delta$ is the relativistic Doppler factor of the emission region, $h$ is the Planck constant, $e$ and $m$ are 
the charge and mass of the electron, c is the velocity of light and B is the source magnetic field responsible for the synchrotron emission. The synchrotron 
power function $f(x)$ is given by
\begin{align}\label{eq:fx}
	f(x) = x\int\limits_x^\infty K_{5/3}(\zeta)\,d\zeta
\end{align}
where, $K_{5/3}$ is the modified Bessel function of order $5/3$. Introducing
\begin{align}\label{eq:alpha} 
	\alpha=\left(\frac{1+z}{\delta}\right)\frac{16m^2c^3}{3heB}
\end{align}
From equation (\ref{eq:synf}) we obtain,
\begin{align}\label{eq:synf2}
	F_\epsilon \propto \frac{1}{\sqrt{\epsilon}}\int\limits_0^\infty f(x)\,{\rm exp}\left[-\frac{\psi}{\kappa} \left( \frac{\epsilon}{mc^2x} \right)^{\kappa/2} \right] \frac{dx}{ \sqrt{x}}
\end{align}
where,
\begin{align}\label{eq:psikappa}
	\psi =\xi_0 \left(\frac{\alpha m^2 c^4}{E_0^2}\right)^{\kappa/2}
\end{align}
Equation (\ref{eq:synf2}) describes the observed synchrotron spectrum in terms of 
two parameters, $\psi$ and $\kappa$. It will be evident from the next section that the parameter $\kappa$ 
governs the spectral curvature and $\psi$ determines the spectral index at a given energy.

\subsection{Synchrotron Spectral Slope}\label{sec:index}

The synchrotron spectrum described by equation (\ref{eq:synf2}) \linebreak  deviates significantly from a power-law and hence
the spectral slope will be a function of the photon energy $\epsilon$. From equation (\ref{eq:eq2}), the slope
of the accelerated particle distribution in logarithmic scale will be 
\begin{align}
\label{eq:index1}
\frac{d\,\rm{ln}\,N_a(E)}{d\,lnE} =-\left(1+\frac{t_{\rm acc}}{t_{\rm esc}}\right) 
\end{align}
Since, $N_c(E) \propto N_a(E)/E$ the slope of the electron distribution responsible for the radiation in 
logarithmic scale will be
\begin{align}
\label{eq:A2}
\frac{d\,\rm{ln}\,N_c}{d\,\rm{ln} \,E} = -\left(2+\frac{t_{\rm acc}}{t_{\rm esc}}\right)
\end{align}

Considering a $\delta$-function approximation for the single particle emissivity, the 
synchrotron emissivity due to the electron distribution $N_c$ will be \citep{Finke+2008,BD2010,Sahayanathan+2012}
\begin{align}\label{jsyn}
	j_{syn}(\epsilon) \propto \epsilon^{1/2} N_c\left(\sqrt{\frac{\epsilon}{\epsilon_L}}\right) 
\end{align}
the synchrotron spectral slope at the photon energy $\epsilon$ will then be,
\begin{align}\label{eq:ind}
	\frac{d\,\rm{ln} j_{syn}}{d\,\rm{ln} \epsilon} & \approx -\frac{1}{2}\left[1+\xi_0\left(\frac{E}{E_0}\right)^\kappa\right]
\end{align}
Where, $E$ is the energy of electron responsible for the emission at photon energy $\epsilon$. 
Since $\epsilon \propto E^2$, the spectral slope  in $\epsilon j_{syn}$ will be,
\begin{align}\label{eq:curv1}
	\frac{d\,\rm{ln} (\epsilon j_{syn})}{d\,\rm{ln} \epsilon}& \approx \frac{1}{2}\left[1-\xi_0\left(\frac{\epsilon}{\epsilon_0}\right)^{\kappa/2}\right]
\end{align}
Here, $\epsilon_0$ is the energy of the photon emitted by the electron of energy $E_0$. 
If we approximate the function $f(x)$ in equation (\ref{eq:synf2}) by a $\delta$-function peaking at 
$\approx 0.29$ \citep{Rybicki_Lightman}, the synchrotron spectral slope in terms of the parameters $\psi$ and $\kappa$ 
can be obtained as
\begin{align}\label{eq:curv2}
	\frac{d\,\rm{ln} (\epsilon F_{\epsilon})}{d\,\rm{ln} \epsilon}& \approx \frac{1}{2}\left[1- \psi\left(\frac{\epsilon}{0.29\,mc^2} \right)^{\kappa/2}\right]  
\end{align}

\subsection{Synchrotron spectral peak and curvature}\label{sec:peak_curv}

At the photon energy $\epsilon_p$, where the flux peaks in $\epsilon F_\epsilon$ representation, the spectral slope will be zero.
Hence, from equation (\ref{eq:curv1}) we obtain,
\begin{align}\label{eq:peak}
	\epsilon_p = \epsilon_0 \,\xi_0^{-2/\kappa}
\end{align}
The spectral curvature at the photon energy $\epsilon_p$ will be\footnote{The curvature of a function $y = g(x)$ is defined as
	$H= g''/(1+g'^2)^{1.5}$}
\begin{align}
	H_p &= \left.{\frac{d^2\,\rm{ln} \,\epsilon j_{syn}}{d\,\rm{ln} \epsilon^2}}\right|_{\epsilon=\epsilon_p}\nonumber \\
		&=-\frac{\kappa}{4}
\end{align}
The above relation illustrates the spectral curvature where the synchrotron spectrum peaks, is governed by the parameter 
($\kappa$) which determines the energy dependence of the escape time-scale. Since the synchrotron flux is maximum at $\epsilon_p$,
$\kappa>0$. Using equation (\ref{eq:peak}), $H_p$ can be written in terms of $\epsilon_p$ as 
\begin{align}
	H_p &= \frac{1}{2}\, \frac{\rm{ln} \,\xi_0}{\rm{ln}\,\left(\epsilon_p/\epsilon_0\right)}
\end{align}
Alternatively, using equation (\ref{eq:curv2}), $H_p$ can also be expressed in terms of $\psi$ and $\epsilon_p$ as
\begin{align}\label{eq:H_psi}
	H_p & \approx \frac{1}{2}\, \frac{\rm{ln} \,\psi}{\rm{ln}\,\left(\frac{\epsilon_p}{0.29\,mc^2}\right)} 
\end{align}
Since $H_p<0$, we attain the condition that $\epsilon_p<0.29\,mc^2$ for $\psi>1$ and $\epsilon_p > 0.29\,mc^2$ for $\psi<1$.
In the former case, the curvature increases with the peak energy and it is reverse in case of the latter.
In other words, a positive correlation between $\kappa$ and ${\rm ln}\,\epsilon_p$ demands $\psi>1$ and vice-versa. 
Similar conclusion can also be obtained from equation (\ref{eq:peak}) since $\kappa>0$.

\section{Spectral analysis}\label{sec:analysis}

The X-ray spectrum of the BL Lac object Mkn\,421 is \linebreak  analysed using the EDD model discussed in the previous section. To obtain the 
synchrotron flux, the double integration given by equations (\ref{eq:fx}) and (\ref{eq:synf2}) needs to be evaluated numerically.
However, to improve the computational speed we approximate $f(x)$ as \citep{melrose1980plasma}
\begin{align}
	f(x) = 1.8\,x^{0.3}\,{\rm exp}(-x)
\end{align}
and this reduces the equation (\ref{eq:synf2}) to single integration which is evaluated using 80 point 
Gaussian-quadratures \citep{1992nrfa.book.....P}. 
The spectral shape of the 
EDD model is described by two parameters, namely $\psi$ and $\kappa$. Alternatively, one can choose the fit parameters
to be $\epsilon_p$ and $\kappa$ using equation (\ref{eq:curv2}). Since the curvature at $\epsilon_p$ is $-\kappa/4$, the latter
approach provides a direct analysis of the synchrotron spectral peak and the associated curvature. However, this demands the
observation to be available in the neighbourhood of $\epsilon_p$ for convincing results. 
We developed two separate numerical codes corresponding
to these choices of fit parameters and added them as local models in {\tt XSPEC} \citep{Arnaud96}. Using these models, the X-ray observations 
of Mkn\,421 are fitted adopting a $\chi^2$-minimization technique. 

\begin{figure}
\centering

	\includegraphics[scale=0.8]{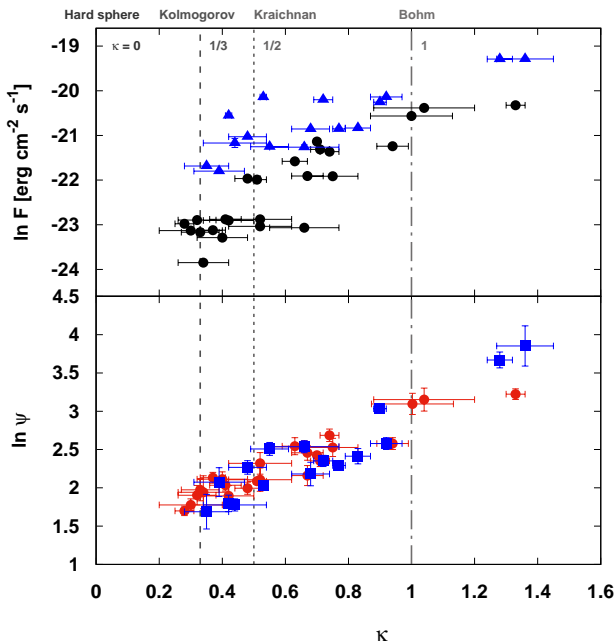}	
	\caption{The correlation plots between the best-fit EDD model parameters and the integrated flux of Mkn\,421 using \emph{Swift}-XRT and \emph{Nu}STAR observations. {\bf Top panel:} Scatter plots between 3--10 keV integrated flux v/s $\kappa$ using \emph{Nu}STAR data alone is represented by black circles and 0.3--10 keV integrated flux v/s $\kappa$ for simultaneous XRT -- \emph{Nu}STAR data is shown by blue triangles. {\bf Bottom panel:} Scatter plot between ln $\psi$ v/s $\kappa$. The red circles indicate the correlation using \emph{Nu}STAR data alone, whereas the blue squares indicate the same for the simultaneous XRT -- \emph{Nu}STAR data. The vertical lines at $\kappa$= 0, 1/3, 1/2 and 1 represent the various type of turbulences corresponding to hard-sphere, Kolmogorov, Kraichnan and Bohm as discussed in \S \ref{sec:turb_index}.}
	 \label{fig1}
\end{figure}

In the present work, we repeated the analysis of 20 \emph{Nu}STAR pointings of the source Mkn\,421 presented in our 
earlier work (Paper-I) using the above two models. In addition, we update our analysis by including 4 new observations
by \emph{Nu}STAR during 2017 and the details are given in Table \ref{table1}. The details about the data reduction procedure are described
in Paper-I. The spectral fit results are summarized in Table \ref{table2}. 
Equation (\ref{eq:psikappa}) implies a linear correlation between ${\rm ln}\,\psi$ and $\kappa$, 
provided $\alpha/E_0^2$ is constant.
In Figure \ref{fig1} (bottom panel), we show the scatter plot between ${\rm ln}\,\psi$ and $\kappa$ and 
the Pearson correlation study resulted in a positive correlation with the correlation coefficient $r=0.95$ 
and null-hypothesis probability $P_p = 1.74\times10^{-12}$. The Spearman
rank correlation study results are, the correlation coefficient $\rho=0.92$ and the null-hypothesis probability $P_s = 1.02\times10^{-10}$.
\emph{Presence of this strong correlation, besides favouring the EDD model, it indicates the quantity $\delta B E_0^2$ is nearly constant for Mkn\;421 independent of the observed flux states}. We will discuss more about this constancy in \S \ref{sec:srcchar}. 


\begin{figure}
\centering

	\includegraphics[scale=0.75]{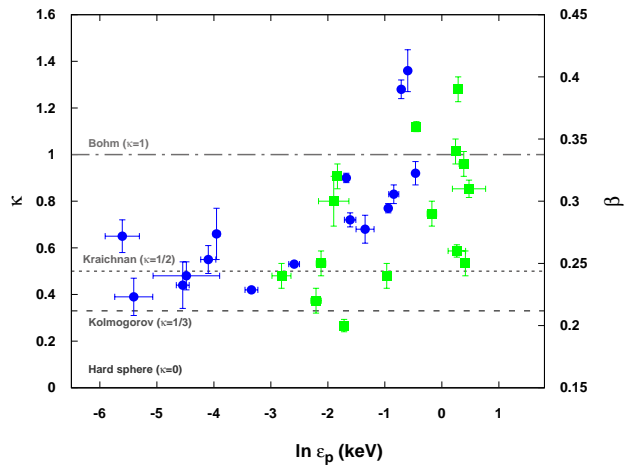}	
	\caption{The scatter plots showing correlation between the curvature parameter and the synchrotron peak using EDD and \emph{eplogpar} model 
fit for 16 different simultaneous XRT -- \emph{Nu}STAR observations. The blue circles represent plot between the parameters $\kappa$ and $\epsilon_{p}$ obtained from the EDD model fit, whereas, the green boxes represent the plot between $\beta$ and $\epsilon_{p,lp}$ obtained from \emph{eplogpar}. The horizontal lines at $\kappa$= 0, 1/3, 1/2 and 1 represent the various type of turbulence as discussed in \S \ref{sec:turb_index}. }
	 \label{fig2}
\end{figure}  


\begin{table*}
\centering
\footnotesize
\caption{Summary of the additional 6 pointings of simultaneous XRT -- \emph{Nu}STAR observations of Mkn\,421.}
\label{table1}
\vspace{0.3cm} 
\begin{tabular}{lcccccr}
\hline \hline
\multicolumn{3}{c}{\emph{Swift}-XRT} && \multicolumn{3}{c}{\emph{Nu}STAR}  \\
\cline{1-3}
\cline{5-7}

Obs. ID& Obs. date $\&$ time &Exposure &&Obs. ID & Obs. date $\&$ time&Exposure \\
     & (dd-mm-yy)          & (sec)   &&        & (dd-mm-yy)         & (sec) \\
\cline{1-3}
\cline{5-7}
00032792001 &2013-04-11 T03:41:30&3489& &60002023025 &2013-04-11 T01:01:07	&57509 	\\
00032792002 &2013-04-14 T00:38:59&6327& &60002023029 &2013-04-13 T21:36:07	&16510    \\
00034228166 &2017-03-27 T22:12:57&1119& &60202048008  &2017-03-27 T21:51:09   &31228    \\
00034228146 &2017-03-01 T00:26:19&85  & &60202048006 & 2017-02-28 T22:11:09   &23906  \\
00034228110 &2017-01-04 T00:06:57& 6021& &60202048002 & 2017-01-03 T23:51:09   &23691	\\
00081926001 &2017-01-31 T23:27:57&1009& &60202048004 & 2017-01-31 T23:46:09  &21564  \\
\hline \hline  
\end{tabular}   \\
\end{table*}


\begin{table*}
\centering
\footnotesize
\caption{The best-fit EDD model parameters for 24 pointings of \emph{Nu}STAR observations.}
\label{table2}
\vspace{0.3cm} 
\begin{tabular}{lccccr}
\hline \hline 
\emph{Nu}STAR Obs. ID  &  $\psi$& $\kappa$ & $\epsilon_{\rm\,p}$ & $\chi_{\rm\,red}^{2}({\rm dof})$& F$_{\rm\,3-10\,keV}$ \\  
\hline 

10002015001&5.47 $\pm$0.34&0.28$\pm$0.03 & $<$2.70$\times 10^{-3}$         &1.09 (731) & 1.05$\pm$0.01 \\  
10002016001&7.64 $\pm$0.57&0.41$\pm$0.05 & (1.25$\pm$0.51)$\times 10^{-2}$ &1.21 (656) & 1.16$\pm$0.01\\                   
60002023006&8.27 $\pm$0.79&0.40$\pm$0.08 & (9.11$\pm$4.32)$\times 10^{-3}$ &0.95 (556) & 0.77$\pm$0.01\\                   
60002023010&10.17$\pm$1.43&0.52$\pm$0.10 & (4.51$\pm$3.20)$\times 10^{-2}$ &1.01 (565) & 0.99$\pm$0.01\\                   
60002023012&6.65 $\pm$0.90&0.42$\pm$0.08 & (4.94$\pm$1.34)$\times 10^{-2}$ &1.01 (582) & 1.13$\pm$0.01\\                   
60002023014&6.97 $\pm$1.50&0.34$\pm$0.08 & $<$4.70$\times 10^{-3}$         &1.05 (414) & 0.44$\pm$0.01\\  
60002023016&6.71 $\pm$0.85&0.32$\pm$0.06 & $<$2.90$\times 10^{-3}$         &1.04 (552) & 1.14$\pm$0.01 \\  
60002023018&7.20 $\pm$1.03&0.33$\pm$0.06 & $<$2.50$\times 10^{-3}$         &0.96 (503) & 0.87$\pm$0.07\\  
60002023020&8.22 $\pm$1.25&0.52$\pm$0.10 & (7.65$\pm$3.01)$\times 10^{-2}$ &0.99 (587) & 1.16$\pm$0.01\\    
60002023022&8.06 $\pm$0.31&0.51$\pm$0.03 & (8.77$\pm$1.10)$\times 10^{-2}$ &1.04 (890) & 2.82$\pm$0.01\\  
60002023024&12.73$\pm$1.41&0.63$\pm$0.04 & (8.03$\pm$3.02)$\times 10^{-2}$ &1.02 (614) & 4.25$\pm$0.03\\  
60002023025&11.31$\pm$0.39&0.70$\pm$0.01 & 0.36                $\pm$ 0.01  &1.04 (1420)& 6.62$\pm$0.01 \\  
60002023026&22.10$\pm$3.05&1.01$\pm$0.13 & 0.77                $\pm$ 0.15  &1.01 (395) &11.72$\pm$0.12 \\
60002023027&23.38$\pm$3.51&1.04$\pm$0.16 & 0.82                $\pm$ 0.03  &0.91 (1011 &14.05$\pm$0.15  \\ 
60002023029&14.67$\pm$1.19&0.74$\pm$0.03 & 0.22                $\pm$ 0.02  &1.09 (908) & 5.27$\pm$0.01 \\ 
60002023031&25.10$\pm$1.75&1.33$\pm$0.03 & 2.50                $\pm$ 0.04  &1.06 (1422)&14.90$\pm$0.18   \\
60002023033&10.50$\pm$0.77&0.71$\pm$0.03 & 0.50                $\pm$ 0.03  &0.95 (1010)& 5.55$\pm$0.01 \\  
60002023035&13.17$\pm$1.02&0.94$\pm$0.05 & 1.54                $\pm$ 0.05  &1.03 (1171)& 5.95$\pm$0.02  \\ 
60002023037&12.64$\pm$0.98&0.66$\pm$0.11 & (8.80$\pm$3.50)$\times 10^{-2}$ &1.01 (564) & 0.96$\pm$0.01\\  
60002023039&5.91 $\pm$0.47&0.30$\pm$0.10 & $<$3.01$\times 10^{-3}$         &0.90 (515) & 0.90$\pm$0.01\\
60202048002$\dagger$&7.35 $\pm$0.58&0.48$\pm$0.04 & 0.13       $\pm$ 0.04  &0.98 (999) & 2.89$\pm$0.02 \\  
60202048004$\dagger$&12.52$\pm$1.49&0.75$\pm$0.08 & 0.37       $\pm$ 0.05  &0.95 (987) & 3.04$\pm$0.01  \\ 
60202048006$\dagger$&11.67$\pm$1.13&0.67$\pm$0.05 & 0.25       $\pm$ 0.04  &1.05 (986) & 3.06$\pm$0.01 \\  
60202048008$\dagger$&8.43 $\pm$0.59&0.37$\pm$0.04 & (1.95$\pm$0.97)$\times 10^{-2}$ &0.95 (704)& 0.91$\pm$0.02 \\
                                                    
\hline \hline 
\end{tabular} \\ 
{\bf Note:} Observation details are given in Paper-I, except for the pointings marked by $\dagger$ symbol which are given in Table \ref{table1}. \linebreak $\epsilon_p$, in units of keV and F$_{3-10\,keV}$, in units of 10$^{-10}$ erg cm$^{-2}$ s$^{-1}$ are calculated from the best-fit EDD model.
\end{table*}


The spectral fit of \emph{Nu}STAR observations of Mkn\,421 using EDD model satisfies the condition $\psi>1$ for all flux states.
Hence, from equations (\ref{eq:H_psi}) and (\ref{eq:peak}), a linear correlation is expected between $\kappa$ and 
${\rm ln}\,\epsilon_p$. To perform the spectral fitting with $\kappa$ and $\epsilon_p$ as free parameters, \emph{Nu}STAR 
observations alone would not be sufficient since the synchrotron spectral peak lies at energies much lower
than its operational energy range. Therefore, we selected the \emph{Nu}STAR observations for which simultaneous observations
by \emph{Swift}-XRT at soft X-ray energies are available. In addition to the 10 observations presented in Paper-I, 
we updated our analysis by including 6 simultaneous observations with 2 during 2013 and 4 during 2017
while maintaining the observation time gap between the X-ray instruments within $\sim$3hrs.
The observational details are given in Table \ref{table1} and the XRT data reduction procedures are 
described in Paper-I. The results of the spectral fit using EDD model with $\kappa$ and $\epsilon_p$ as fit parameters 
are given in Table \ref{table3} and the scatter plot between ${\rm ln}\,\epsilon_p$ and $\kappa$ is shown in  Figure \ref{fig2} (blue circles).
A strong positive correlation is observed between ${\rm ln}\,\epsilon_p$ and $\kappa$ with
$r=0.86$; $P_p = 2.42\times10^{-5}$ and $\rho=0.91$; $P_s = 7.01\times10^{-7}$, and this is again 
consistent with the EDD model. For comparison, we fitted the simultaneous 
XRT -- \emph{Nu}STAR 
observations using \emph{eplogpar} model inbuilt in XSPEC \citep{2007A&A...466..521T,2009A&A...501..879T}. The 
fit results are given in Table \ref{table3} and the scatter plot between the synchrotron peak $\epsilon_p$ 
and the curvature parameter $\beta$ of  
\emph{eplogpar} is shown in Figure \ref{fig2} (green points). We see a trend of $\beta$ increasing with $\epsilon_p$, with marginal correlation 
coefficients of $r=0.56$ ($P_p = 0.02$) and $\rho=0.53$ ($P_s= 0.03$). For the correlation study, we have used only those data points with associated uncertainity less than 20$\%$. This is to assure that results are not biased by the large uncertainities.

We also fitted the simultaneous XRT -- \emph{Nu}STAR 
observations using the EDD model with $\psi$ and $\kappa$ as fit parameters and this reduced the correlation
with $r=0.90$; $P_p = 1.65\times10^{-6}$ and $\rho=0.86$; $P_s = 5.14\times10^{-6}$. In Figure \ref{fig1}, the scatter plot between
the best-fit ${\rm ln}\,\psi$ and $\kappa$ is shown as blue squares for the simultaneous XRT -- \emph{Nu}STAR 
observations. The reduction in the correlation between ${\rm ln}\,\psi$ and $\kappa$ with the inclusion of soft X-ray data implies
the EDD model may not be favourable at low energies.
A careful observation of equations (\ref{eq:tesc}) and (\ref{eq:kin}) may highlight the plausible reason for this. As the 
particle escape rate is an increasing function of particle energy ($\because \kappa>0$), the effect of the EDD model will be 
more prominent at higher particle energy rather than the lower ones. This is also consistent with the case that the
probability to confine electron within the acceleration region decreases with the increase in energy.
Accordingly, the hard X-ray spectra will reflect the EDD model better than the soft X-ray.

\subsection{Turbulence index and $\kappa$}\label{sec:turb_index}
The escape of the electron from the vicinity of the shock is related to the spatial diffusion. In the jet medium,
the scattering of the particle by the magnetohydrodynamic wave associated with the turbulence 
governs the spatial diffusion \citep{Blandford1987PhR}.
We express the turbulent energy spectrum $W(k)$ as,
\begin{align}
W(k) \propto k^{-q}
\end{align}
where, $k$ is the wave number and the range of the turbulent spectral index is $1\le q\le 2$. The lower limit, $q=1$, 
corresponds to Bohm diffusion where the mean free path scales as the gyroradius of the charged particle. On the 
other hand, the upper limit, $q=2$, corresponds to the hard-sphere diffusion approximation where the mean free path
is independent of the energy of the charged particle. The diffusion due to Kolmogorov type turbulence corresponds to 
$q=5/3$, and in case of Kraichnan type $q=3/2$ \citep{2019ApJ...877...71T}. 

The escape time-scale from the acceleration region due to this spatial diffusion can be written as \citep{2002cra..book.....S},
\begin{align}\label{eq:qindex}
	t_{esc} \propto E^{q-2}
\end{align}
Comparing with equation (\ref{eq:tesc}) we find the parameter $\kappa$ as, $\kappa = 2-q$ and the allowed range of $\kappa$
will be $0<\kappa<1$. For Kolmogorov type of turbulence, $\kappa = 1/3$ and in case of Kraichnan type, $\kappa=1/2$.
From Figure \ref{fig1}, its evident that the best-fit $\kappa$ largely satisfies within the allowed range.
Most of the fitted $\kappa$ values are consistent with Kolmogorov and Kraichnan case with some falling in between 
the Kraichnan and Bohm limit. Incidentally, no observations favoured the hard-sphere limit in the underlying assumptions of EDD model. 
In addition, two points fall beyond the Bohm limit are unphysical and highlight the limitation of EDD model. 
We further identify the probable association of the flux level with the type of turbulence by performing the correlation analysis between the integrated 
flux ($F$) and $\kappa$.
In Figure \ref{fig1} (top), we plot the ${\rm ln} F$ against the best-fit $\kappa$ for the observations considered in 
the present work.
A strong positive correlation between ${\rm ln} F$ and $\kappa$ is witnessed with $r = 0.91$; $P_p = 6.35\times10^{-10}$ and 
$\rho = 0.85$; $P_s = 1.36\times10^{-7}$ when the spectral fitting is performed on \emph{Nu}STAR observation alone.
However, the correlation reduced for the case of simultaneous XRT -- \emph{Nu}STAR 
spectral fit, and this again hints the EDD model may not be favourable at low energies. From the positive 
correlation between ${\rm ln} F$ and $\kappa$, one can speculate 
that the magnetohydrodynamic turbulence in the jet of Mkn\,421 tend to approach the Bohm limit during high flux.

\section{Characteristics of fit Parameters}\label{sec:srcchar}

The strong linear correlation between ${\rm ln}\,\psi$ and $\kappa$, put together with equation (\ref{eq:psikappa}),
indicates the quantity $\delta B E_0^2$ is nearly constant for Mkn\;421, independent of observed flux states.
This quantity can be estimated from the slope of the linear fit to ${\rm ln}\,\psi$ vs $\kappa$ plot. 
For the case of \emph{Nu}STAR observation, a least-square fit to these quantities, taking into account the 
uncertainties \citep{1992nrfa.book.....P}, resulted in 
slope $a=1.44\pm0.08$ and the y-intercept $b=1.41\pm0.05$ with betterness of fit $q$-value $=0.26$.  
Likewise, the least square fit for the case of simultaneous
XRT-\emph{Nu}STAR analysis, resulted in slope $a=1.96\pm0.18$ and the 
y-intercept $b=1.05\pm0.14$ with betterness of fit $q$-value $=0.22$. 

The slope of the best-fit line can be used to estimate the product $\delta B$,
provided if $E_0$ can be eliminated. From equation (\ref{eq:psikappa}), the slope of the linear relation between
ln\,$\psi$ and $\kappa$ will be,
\begin{align}\label{eq:slope}
	a \approx \frac{1}{2} {\rm ln} \left[10^{19}\,\left(\frac{1+z}{\delta B}\right)\,\left(\frac{E_0}{\rm keV}\right)^{-2} \right]
\end{align}
and the y-intercept
\begin{align}\label{eq:yint}
	b = {\ln}\, \xi_0
\end{align}
Since in $\epsilon F_\epsilon$ representation the spectral slope at $\epsilon_p$ is zero, from equation (\ref{eq:ind})
and (\ref{eq:yint}) we obtain,
\begin{align}\label{eq:e0}
	E_0=E_p\,{\rm exp}\,\left(\frac{b}{\kappa}\right)
\end{align}
where, $E_p$ is the electron energy responsible for the emission at $\epsilon_p$. Under synchrotron theory, $E_p$ 
and $\epsilon_p$ are related as \citep{Rybicki_Lightman,1991par..book.....S}
\begin{align}\label{eq:synpk}
	\epsilon_p \approx \left(\frac{\delta}{1+z}\right) \left(\frac{E_p}{mc^2}\right)^2 \left(\frac{eB}{2\pi mc}\right) 
\end{align}
If $\epsilon_{ic}$ is the energy at which the $\gamma$-ray component of the SED peaks then assuming a 
synchrotron self Compton origin we can express \citep{Rybicki_Lightman}
\begin{align}\label{eq:icpk}
	\epsilon_{ic} \approx \left(\frac{\delta}{1+z}\right) \left(\frac{E_p}{mc^2}\right)^4 \left(\frac{eB}{2\pi mc}\right) 
\end{align}
From equations (\ref{eq:synpk}) and (\ref{eq:icpk})
\begin{align}\label{eq:ep}
	E_p=mc^2\sqrt{\frac{\epsilon_{ic}}{\epsilon_p}}
\end{align}
Finally, using equations (\ref{eq:slope}), (\ref{eq:e0}) and  (\ref{eq:ep}), we obtain the product of $\delta B$ in terms of the best-fit values of the linear relation between ln\,$\psi$ and $\kappa$ as,
\begin{align}\label{eq:bd1}
	(\delta B)_{\rm\,EDD}\approx 3.75\times 10^{10}(1+z)\left(\frac{\epsilon_p}{\rm keV}\right)\left(\frac{\epsilon_{ic}}{{\rm MeV}}\right)^{-1}{\rm exp}\left[-2 \left( a+\frac{b}{\kappa}\right)\right]
\end{align}

Note that the above equation highlights \emph{the association of the energy dependence of the particle escape time-scale 
with the source magnetic field and the Doppler factor of the emission region.}

An alternate expression of $\delta B$ can be obtained in terms of the SED peak energies using equations (\ref{eq:synpk}), (\ref{eq:icpk}) and (\ref{eq:ep}) as, 
\begin{align} \label{eq:bd2}
	(\delta B)_{\rm\,peak}\approx 8.6\times 10^{7}(1+z)\left(\frac{\epsilon_p}{{\rm keV}}\right)^2 \left(\frac{\epsilon_{ic}}{{\rm MeV}}\right)^{-1}
\end{align}
If the energy dependence of the escape time-scale is related to the product of the source magnetic field and the Doppler factor
then one would expect the $\delta B$ estimates using equations (\ref{eq:bd1}) and (\ref{eq:bd2}) to be equal. However, to verify 
this, prior knowledge of the SED peak energies is required. For $\epsilon_p$, besides using the values obtained from EDD model, 
we also fit the simultaneous XRT -- \emph{Nu}STAR observations using \emph{eplogpar} model as an alternate estimate as discussed in \S \ref{sec:analysis}. 
The synchrotron peak values corresponding to the latter are denoted as $\epsilon_{p,lp}$ and given in Table \ref{table3}.
The non-availability of simultaneous $\gamma$-ray observations do not allow us to identify $\epsilon_{ic}$ and 
hence, we employ an indirect method
to estimate $\epsilon_{ic}$ from the archival peak energies.  

\subsection{Compton Spectral Peak ($\epsilon_{ic}$)}\label{sec:ic_peak}

\begin{figure}
\centering

	\includegraphics[scale=0.75]{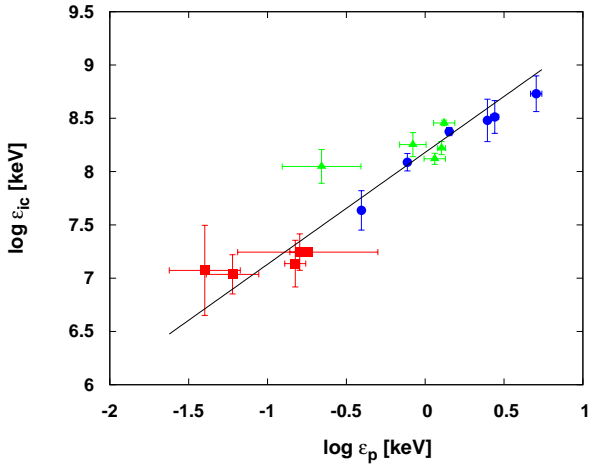}	 
	\caption{The correlation plot between the quasi-simultaneous archival $\epsilon_p$ and $\epsilon_{ic}$. The red boxes are for MJD 56307 -- 56362, where the $\epsilon_p$ values were estimated using \emph{eplogpar} fit of XRT data and corresponding $\epsilon_{ic}$ were extracted using the spectral information given in ~\protect\cite{mislav2016}. The blue circles were extracted using the spectral information given in ~\protect\cite{bartoli2016} for MJD 55144 -- 56106. The green triangles are for MJD 55240 -- 55246 where the $\epsilon_p$ values were estimated using \emph{eplogpar} fit of XRT data and the corresponding $\epsilon_{ic}$ extracted from ~\protect\cite{Kapanadze+2016}. The black line represents the best-fit straight line.}
	 \label{fig3}
\end{figure}

For an appropriate estimate of $\epsilon_{ic}$, we study the correlation between the archival SED peaks 
($\epsilon_p$ and $\epsilon_{ic}$) extracted from simultaneous/near-simultaneous multi wavelength observations 
during different flux states \citep{mislav2016, bartoli2016,Kapanadze+2016}. The $\epsilon_p$ during the 
period MJD 55144-56106 is directly obtained from \cite{bartoli2016}; whereas for MJD 56307-56362, we
fitted the XRT observations using \emph{eplogpar} model to estimate $\epsilon_p$. 
The $\epsilon_{ic}$ during both these periods are estimated by extrapolating the power-law $\gamma$-ray 
spectra corresponding to \emph{Fermi} and MAGIC/VERITAS/ARGO-YBJ observations \citep{bartoli2016,mislav2016}. 
In Figure \ref{fig3}, these estimates are represented
as blue solid circles for MJD 55144-56106 and red solid squares for MJD  56307-56362. 
For the period MJD 56393-56397, we estimate $\epsilon_p$ from log-parabola spectral analysis of simultaneous XRT--\emph{Nu}STAR observations and 
the corresponding $\epsilon_{ic}$ values are obtained from the estimation using log-parabola spectral fit of MAGIC observations, performed 
by \cite{Kapanadze+2016} and \cite{Preziuso+2013}. These estimates are shown as green solid triangles in Figure \ref{fig3}.

\begin{table*}
\centering
\scriptsize
\caption{The best-fit EDD model parameters and the estimated quantities for the simultaneous XRT -- \emph{Nu}STAR observations.}
\label{table3}
\vspace{0.3cm} 
\begin{tabular}{lcccccccccr}
\hline \hline 
\multicolumn{2}{c}{Obs. ID } &\multicolumn{4}{c}{EDD model} & $\epsilon_{\rm\,ic}$ & $\epsilon_{\rm\,p,lp}$ & $\epsilon_{\rm\,ic,lp}$ &F$_{\rm\,0.3-10\,keV}$  \\
\cline{3-6} 
XRT &\emph{Nu}STAR   &  $\psi$& $\kappa$ & $\epsilon_{\rm\,p}$ (keV) & $\chi_{\rm red}^{2}({\rm dof})$ &(GeV)& (keV)& (GeV) & \\  
 $[1]$ & [2] &  [3]& [4] & [5] & [6] & [7] & [8] & [9] & [10] \\ 

\hline

00080050006&60002023014  &7.96 $\pm$ 1.50&0.39$\pm$0.08&(4.49$\pm$1.50)$\times 10^{-3}$&1.20 (779) & 1.11$\pm$0.33&0.06$\pm$0.01 & 11.49$\pm$1.72 & 3.41$\pm$0.03\\
00080050007&60002023016  &9.65 $\pm$ 0.85&0.48$\pm$0.06&(1.13$\pm$0.66)$\times 10^{-2}$&1.05 (760) & 2.55$\pm$1.34&0.15$\pm$0.04 & 26.21$\pm$6.29 & 7.37$\pm$0.17\\
00080050011&60002023018  &12.28$\pm$ 1.03&0.55$\pm$0.06&(1.66$\pm$0.22)$\times 10^{-2}$&1.008(921) & 3.61$\pm$0.43&0.16$\pm$0.01 & 27.77$\pm$1.56 & 5.88$\pm$0.03\\
00032792001$\dagger$&60002023025$\dagger$  &6.03 $\pm$ 0.65&0.42$\pm$0.01&(3.54$\pm$0.40)$\times 10^{-2}$&1.48 (1836)& 7.14$\pm$0.72&0.18$\pm$0.01 & 30.88$\pm$1.54 &11.83$\pm$0.08\\
00080050016&60002023025&7.61 $\pm$ 0.28&0.53$\pm$0.01&(7.41$\pm$0.79)$\times 10^{-2}$&1.29 (1780)&14.02$\pm$1.19&0.38$\pm$0.02 & 60.50$\pm$2.86 &17.88$\pm$0.19 \\
00080050019&60002023026  &47.10$\pm$12.35&1.36$\pm$0.09&0.55                $\pm$ 0.02 &1.20 (903) &84.39$\pm$3.03&1.27$\pm$0.04 &179.23$\pm$5.08 &42.02$\pm$1.04 \\
00080050019&60002023027  &39.24$\pm$ 4.05&1.28$\pm$0.04&0.49                $\pm$ 0.02 &1.28 (1519)&76.06$\pm$2.79&1.33$\pm$0.03 &186.83$\pm$3.79 &41.86$\pm$0.27 \\
00032792002$\dagger$&60002023029$\dagger$&20.75$\pm$ 0.99&0.90$\pm$0.02&0.18                $\pm$ 0.01 &1.27 (1449)&31.96$\pm$0.95&0.64$\pm$0.01 & 96.73$\pm$1.36 &15.95$\pm$0.07 \\
00035014062&60002023033  &10.50$\pm$ 0.77&0.72$\pm$0.03&0.20                $\pm$ 0.02 &0.98 (1309)&33.95$\pm$3.05&0.84$\pm$0.07 &123.55$\pm$9.26 &16.85$\pm$0.23 \\
00035014065&60002023035  &13.17$\pm$ 1.02&0.92$\pm$0.05&0.63                $\pm$ 0.01 &1.29 (1733)&95.36$\pm$2.45&1.51$\pm$0.03 &209.44$\pm$3.74 &17.93$\pm$0.12 \\
00035014066&60002023037  &12.64$\pm$ 0.98&0.66$\pm$0.11&(1.92$\pm$0.11)$\times 10^{-2}$&1.10 (996) & 4.12$\pm$0.21&0.11$\pm$0.01 & 19.82$\pm$1.62 &5.83$\pm$0.31 \\
00035014067&60002023039  &5.91 $\pm$ 0.47&0.44$\pm$0.10&(1.06$\pm$0.12)$\times 10^{-2}$&1.05 (931) & 2.41$\pm$0.24&0.12$\pm$0.01 & 21.44$\pm$1.60 &6.40$\pm$0.64 \\
00034228166$\dagger$&60202048008$\dagger$&5.41 $\pm$ 0.38&0.35$\pm$0.07&(3.67$\pm$1.10)$\times10^{-3}$&1.02 (879)&0.92$\pm$0.25&0.07$\pm$0.06&12.87$\pm$1.03&3.82$\pm$0.09\\
00034228146$\dagger$&60202048006$\dagger$&8.85 $\pm$ 1.36&0.68$\pm$0.06&0.26 $\pm$ 0.04&1.04 (1012)&43.05$\pm$6.10&1.47$\pm$0.12&204.44$\pm$15.02&8.75$\pm$0.15\\
00034228110$\dagger$&60202048002$\dagger$&9.90 $\pm$ 0.59&0.77$\pm$0.02&0.39$\pm$ 0.01&1.16 (1483)&61.93$\pm$2.57&1.30$\pm$0.20&183.04$\pm$25.34&8.79$\pm$0.21 \\
00081926001$\dagger$&60202048004$\dagger$&11.17$\pm$ 1.12&0.83$\pm$0.04&0.43$\pm$ 0.03&1.005 (1266)&67.62$\pm$5.37&1.61$\pm$0.47&221.89$\pm$58.29&8.96$\pm$0.13\\
\hline \hline 
\end{tabular} \\
{\bf Note:} {\tt Columns}:- [3],[4]$\&$[5]: best-fit EDD model parameters; [7]: Compton peaks estimated from $\epsilon_p$ using equation (\ref{eq:peaks}); [8]$\&$[9]: synchrotron peaks estimated using \emph{eplogpar} model fit and the corresponding Compton peaks estimated from $\epsilon_{p,lp}$ using equation (\ref{eq:peaks}); [10]: 0.3 --10 keV integrated flux in units of 10$^{-10}$ erg cm$^{-2}$ s$^{-1}$, calculated from the best-fit EDD model. Observation details of XRT are given in Paper-I, except for the pointings marked by $\dagger$ symbol which are given in Table \ref{table1}.
\end{table*}

A Spearman rank correlation analysis between this archival $\epsilon_p$ and $\epsilon_{ic}$ show a strong positive 
correlation with $\rho= 0.89$ and $P_s=3.85\times10^{-6}$. Additionally, the Pearson analysis supported a linear correlation with $r=0.91$
and $P_p=1.76\times10^{-6}$ and consequently, we perform a linear regression analysis between $\epsilon_p$ and $\epsilon_{ic}$.
A least-square fit between these quantities gave us the relation,
\begin{align}\label{eq:peaks}
	{\rm log}\, \epsilon_{ic} = (0.90\pm 0.08) \,{\rm log}\, \epsilon_p + (8.16\pm 0.05)
\end{align}
with $q$-value $=0.21$. We used equation (\ref{eq:peaks}) to infer $\epsilon_{ic}$ corresponding to the synchrotron peak 
estimated from simultaneous XRT--\emph{Nu}STAR observations used in this work. 
The inferred values of $\epsilon_{ic}$ are given in column 7 and 9 of Table \ref{table3}. It should be noted that the best-fit line obtained using 
a least-square fit shown in Figure \ref{fig3} involves data points with $\epsilon_p < $ 0.5 keV, which is beyond the observational energy range of XRT. 
However, even considering only the data point with $\epsilon_p \geq $ 0.5 keV the least-square fit only shows a marginal change with best-fit values, 
slope = 0.92 $\pm$ 0.15 and the y-intercept = 8.21 $\pm$ 0.03.

\begin{figure}
\centering

	\includegraphics[scale=0.75]{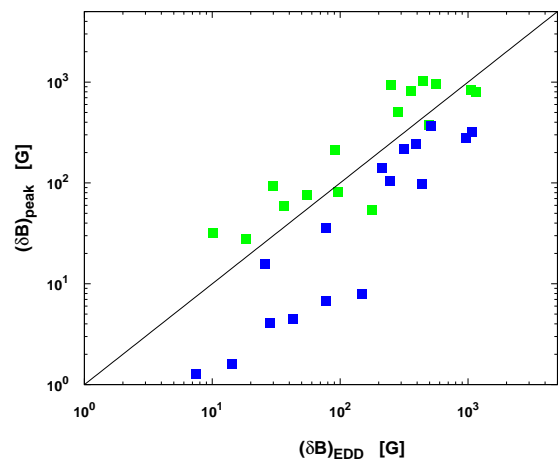}	
	\caption{The plot between $(\delta B)_{\rm\,EDD}$ and $(\delta B)_{\rm\,peaks}$ obtained using equations (\ref{eq:bd1}) and  (\ref{eq:bd2}) for the simultaneous XRT--\emph{Nu}STAR data. The blue boxes represent ($\delta B$) estimation where $\epsilon_p$ values were determined from the EDD model fit. On the other hand, the green boxes indicate these estimations where $\epsilon_{p,lp}$ values were estimated using \emph{eplogpar} fit (Table \ref{table4}). The black line represents the identity line.}
	 \label{fig4}
\end{figure}

The knowledge of $\epsilon_p$ and $\epsilon_{ic}$ allow us to evaluate $\delta B$ described by equations (\ref{eq:bd1})
and (\ref{eq:bd2}) and these estimates are given in Table \ref{table4}. In figure \ref{fig4}, we plot these $\delta B$ values 
along with the identity line. The $\delta B$ values are reasonably close to the identity lines
particularly when $\epsilon_p$ is estimated using \emph{eplogpar} model. This may again be associated with the 
weakness of the EDD model to explain the soft X-ray spectrum. We also note a 
large variation in $\delta B$ and this may probably reflect the extreme variability behaviour of the source. Absence of 
a definite consensus on the range of these parameters may support this argument 
\citep{1995MNRAS.273..583D,Fossati+2008,Mankuzhiyil_2011,bartoli2016,Kapanadze+2016,mislav2016,Banerjee+2019}. 
Alternatively, this variation in $\delta B$ may be associated with the inaccurate estimation of the spectral peak energies or the combination
of both. Nevertheless, this result open up the avenue where the microscopic description about the emitting electron
distribution is related to the parameters governing the overall SED of the source.

\section{Discussion and Summary} \label{sec:summary}

The X-ray spectral curvature of Mkn\,421 is interpreted as the result of energy-dependent electron diffusion from the 
acceleration region (EDD model). 
Unlike the log-parabola model, the EDD model parameters, $\psi$ and $\kappa$ show strong correlation. In this work, we perform a careful 
investigation of the EDD model and the observed correlations. We identify the photon energy 
at which the synchrotron spectral component peaks can correlate/anti-correlate with the spectral 
curvature at this energy. The sense of the correlation is determined by the parameter $\psi$, when $\psi>1$, the correlation is positive and in case of 
$\psi<1$ the quantities are anti correlated. The spectral fitting of Mkn\,421 for the observations during the period MJD 56115-56401 and 57756-57839 indicate
 $\psi$>1 and this suggests the correlation to be positive. This is further verified by performing correlation analysis between the parameters 
$\kappa$ and $\epsilon_p$ using the EDD model with $\epsilon_p$ as a fit parameter. \\

Addition of soft X-ray information from XRT marginally reduces the correlation between the fit parameters and also shows relatively poor fits to some of the combined XRT -- \emph{Nu}STAR spectra, suggesting the EDD model to be better agreeable at hard X-rays. 
A possible reason can be, the effect of escape time-scale is more prominent at higher 
electron energies. Another reason may be related to the simplicity of the model and the underlying assumptions. A detailed description of the model including stochastic acceleration(Fermi second-order process) and/or shock-drift acceleration may have the capability to explain the spectral behaviour over a broad energy range. For instance, \citep{Katarzynski+2006} have studied the evolution of the particle spectrum considering the combination of shock and stochastic acceleration mechanisms. Similarly, \cite{2019ApJ...887..133B} have illustrated the radiative signatures by considering time-dependent evolution of particles due to 
a combination of diffusive shock acceleration and shock drift acceleration. We have assumed a monoenergetic injection of particles into the shock which is then 
accelerated to relativistic energies; however, a detailed description may involve an energy-dependent particle injection 
\citep{2012ApJ...745...63S,Baring+2017}. Nevertheless, an exact description may increase the number of free parameters and constraining them using statistical 
fitting procedure over a narrow energy band may be impossible or misleading.

\begin{table*}
\centering
\footnotesize
\caption{Estimations of $\delta B$ from equations (\ref{eq:bd1}) and (\ref{eq:bd2}) using simultaneous XRT--\emph{Nu}STAR observations.}
\label{table4}
\vspace{0.3cm} 
\begin{tabular}{lcccccr}
\hline \hline 
XRT Obs. ID&\emph{Nu}STAR Obs. ID & ($\delta B)_{\rm\,EDD}$ &  ($\delta B)_{\rm\,peak}$ & ($\delta B)_{\rm\,EDD,lp}$ &  ($\delta B)_{\rm\,peak,lp}$ \\  
 $[1]$ & [2] &  [3]& [4] &  [5]& [6]\\
\hline

00080050006&60002023014  &  14.17&  1.60 &   18.37& 27.77\\
00080050007&60002023016  &  42.67&  4.42 &   55.26& 76.11\\
00080050011&60002023018  &  77.38&  6.75 &   97.06& 81.71\\
00032792001&60002023025  &  25.60& 15.54 &   30.12& 93.01\\
00080050016&60002023025  &  77.89& 35.45 &   91.63&211.60 \\
00080050019&60002023026  &1067.30&317.80 & 1160.47&797.88 \\
00080050019&60002023027  & 957.98&279.88 & 1058.58&839.44 \\
00032792002&60002023029  & 435.19& 97.05 &  492.17&375.45 \\
00035014062&60002023033  & 244.49&104.44 &  282.22&506.36 \\
00035014065&60002023035  & 516.96&369.08 &  564.19&965.23 \\
00035014066&60002023037  & 148.36&  7.93 &  176.66& 54.11 \\
00035014067&60002023039  &  28.48&  4.12 &   36.30& 59.54 \\
00034228166&60202048008&  7.50&  1.28& 10.05&	 31.93\\
00034228146&60202048006&211.42&139.38&251.41& 937.14\\
00034228110&60202048002&315.88&217.73&356.29& 818.64 \\
00081926001&60202048004&388.49&242.42&443.32&1035.77 \\
\hline \hline 
\end{tabular} \\
{\bf Note:} $\delta B$ is in the units of Gauss. {\tt Columns} [3]$\&$[4]: ($\delta B$) estimates involve $\epsilon_{ic}$ values evaluated from the best-fit $\epsilon_p$ values \linebreak using EDD model. $[5]\&$[6]: ($\delta B$) estimates involve $\epsilon_{ic}$ values evaluated from best-fit $\epsilon_{p,lp}$ values using \emph{eplogpar} model.  
\end{table*}

The energy dependence of the escape time-scale of the EDD model can be associated with the magnetohydrodynamic turbulence
responsible for spatial diffusion of the particle. For the case of Mkn\,421, we found this energy dependence satisfies well
within the constraints imposed by the hard-sphere and Bohm diffusion limits. The observed strong correlation between the integrated flux and the energy 
index of the escape time-scale signifies the high flux states may be associated with the turbulence shifting from
Kolmogorov/Kraichnan type to Bohm limit. Incidentally, the spectral fitting of the curved X-ray spectrum by EDD model do not favour hard-sphere type 
turbulence. Under hard-sphere approximation, the parameter $\kappa$ will be zero (equation \ref{eq:qindex}) and from equation (\ref{eq:eq2}) it is evident 
EDD model predicts a 
power-law electron distribution. Since the X-ray spectra of Mkn\,421 is significantly curved, the EDD model does not favour hard-sphere type turbulence. 
However, this conclusion should be dealt with caution due to varying underlying assumptions of the EDD model. Presence of two data points beyond Bohm limit 
(Figure \ref{fig1} $\&$ \ref{fig2}) further indicate the limitation of EDD model. It should also be noted that studies on X-ray spectra of Mkn\,421 using stochastic 
acceleration 
scenario have supported hard-sphere type turbulence \citep{2011ApJ...739...66T, Asano2018}. By means of a Monte Carlo description, \cite{2011ApJ...739...66T} 
have demonstrated the transition from Kraichnan turbulence into the hard-sphere type. Similar results are also reported by several authors for Mkn\,421 for 
different observational epochs \citep{2018ApJ...854...66K,2018ApJ...858...68K,2020ApJS..247...27K}.  \\
 
A least square fit to the EDD model parameters \linebreak allow us to derive an expression for the product of the source magnetic 
field and the jet Doppler factor, in terms of synchrotron and the Compton spectral peaks. 
The peak energy of the synchrotron component is obtained by fitting the simultaneous XRT--\emph{Nu}STAR observations by EDD and \emph{eplogpar} models; whereas, for the peak energy of the Compton spectral component an indirect approach involving archival 
observations is employed. The product of the source magnetic field and the jet Doppler factor estimated through this
approach matches reasonably well with the estimates attained through synchrotron and synchrotron self Compton theory.
This possibly highlights the link between the shape of electron distribution with the parameters deciding the broadband
spectrum of the source. The estimated product of the source magnetic field and the jet Doppler factor varies over a 
wide range $\sim$7 -- 1160. This may either reflect the extreme variable behaviour of the source or may be associated with the 
inaccurate estimation of peak photon energies. 

The EDD model can be further refined to represent a more realistic scenario by incorporating the exact form of the 
momentum diffusion and/or an appropriate energy dependence of the acceleration
time scale. Future works, the inclusion of other available observational signatures will allow us to relax various assumptions and let us step towards a 
global picture of particle acceleration and diffusion in blazar jets. Such changes also have the potential to refine/reject the scenario portrayed by the EDD model. 

\subsection*{Acknowledgements} 
The authors thank the anonymous referee for useful suggestions and comments. This research has made use of data, software,
and/or web tools obtained from NASAs High Energy Astrophysics Science Archive Research Center (HEASARC), a service of the Goddard Space Flight Center 
and the Smithsonian Astrophysical Observatory. RG and PG would like to thank
the ISRO, Department of Space, India for the grant under $"$Space Science Promotion$"$. PG would like to thank IUCAA, BARC and HBCSE for their hospitality during the visit. RG thanks the IUCAA associateship programme.  

\subsection*{DATA AVAILABILITY}

The data underlying this article are available in the HEASARC archive at \url{https://heasarc.gsfc.nasa.gov/}.


\bibliographystyle{mnras}
\bibliography{references} 
\bsp	
\label{lastpage}
\end{document}